\documentclass[manuscript]{raa}            % referee version: for submission

\usepackage{graphicx,times,epsf}             %for PS/EPS graphics inclusion, new

\newcommand{\ergcms}{\mathrm{erg\,cm^{-2}\,s^{-1}}}
\newcommand{\ff}{f_X/f_{opt}}

\begin{document}

   \title{Sharp \emph{Chandra} View of \emph{ROSAT} All-Sky Survey Bright Sources
%\,$^*$
%\footnotetext{$*$ Supported by the National Natural Science Foundation of China.}
}
   \subtitle{I. Improvement of Positional Accuracy}

   \volnopage{Vol.0 (2016) No.0, 000--000}      %%preserved for Editor. DOn't remove!
   \setcounter{page}{1}          %%starting page, preserved for Editor. DOn't remove!

   \author{Shuang Gao
      \inst{1}
   \and Song Wang
      \inst{2}
   \and Jifeng Liu
      \inst{2,3}
   }
%% Here is an example of three authors come from different institutes.
%% For single author or all the authors from an institute, use "\inst{}" only

   \institute{Department of Astronomy, Beijing Normal University, Beijing 100875, China; {\it sgao@bnu.edu.cn}\\
%% Please give the E-mail address of the author, to whom future correspondence and
%% offprint requests will be sent.
        \and
             Key Laboratory of Optical Astronomy, National Astronomical Observatories, Chinese Academy of Sciences, Beijing 100012, China\\
        \and
             College of Astronomy and Space Sciences, University of Chinese Academy of Sciences, Beijing 100049, China\\
   }

   \date{Received~~2016 month day; accepted~~2016~~month day}

\abstract{The \emph{ROSAT} All-Sky Survey (RASS) represents one of the most
   complete and sensitive soft X-ray all-sky surveys to date. However, the
   deficient positional accuracy of the RASS Bright Source Catalog
   (BSC) and subsequent lack of firm optical identifications affect
   the multi-wavelength studies of X-ray sources. The widely used
   positional errors $\sigma_{pos}$ based on the Tycho Stars Catalog
   (Tycho-1) have previously been applied for identifying objects in
   the optical band. The considerably sharper \emph{Chandra} view
   covers a fraction of RASS sources, whose $\sigma_{pos}$ could be
   improved by utilizing the sub-arcsec positional accuracy of
   \emph{Chandra} observations. We cross-match X-ray objects between the
   BSC and \emph{Chandra} sources extracted from the Advanced CCD Imaging
   Spectrometer (ACIS) archival observations. A combined counterparts
   list (BSCxACIS) with \emph{Chandra} spatial positions weighted by
   the X-ray flux of multi-counterparts is employed to evaluate and
   improve the former identifications of BSC with the other
   surveys. Based on these identification evaluations, we suggest that
   the point-likeness of BSC sources and INS (isolated neutron stars) candidates should be carefully reconsidered.
}

   \authorrunning{S, Gao, S. Wang \& J. Liu }            %author_head in even pages
   \titlerunning{Improvement of Positional Accuracy of ROSAT}  % title_head in odd pages

   \maketitle
%% The author head (on even pages) and the title head (on odd pages) will be
%% automatically extracted from \author{} and \title{}. Whenever the title is too long,
%% you will be asked to supply a shorter one by inserting either \authorrunning{} or
%% \titlerunning{} before \maketitle. Anyway, you can specify your own heads.
%%
%%
%% Note: In the following text body of your manuscript, please note several differences from
%%       other major journals:
%% (1) \subsection{Please Capitalize the First Letter of Each Notional Word in Subsection Title}
%% (2) Please Capitalize the First Letter of Each Notional Word in all tables' captions

%
%________________________________________________ sections below
%
\section{Introduction}           %% first-level sections will be auto-capitalized
\label{sect:intro}

The \emph{ROSAT} All-Sky Survey (RASS), the vast majority of which was conducted during the first half-year (1990/1991) of the \emph{ROSAT} mission (Tr\"{u}mper 1983), currently represents one of the most complete and sensitive soft X-ray all-sky surveys. RASS covers $92\%$ of the sky and is 20-fold more sensitive than any previous X-ray survey, with a brightness limit of $0.1$ \emph{ROSAT} position-sensitive proportional counter (PSPC) cts/s (i.e., $\sim1\times10^{-12} \,\ergcms$) in the $0.1-2.4$ keV energy band. The Bright Source Catalog (BSC, Voges et al. 1999) contains the brightest 18,811 sources from RASS-BSC across the majority of the sky. At present, the \emph{Chandra} and XMM-Newton X-ray observatories have improved the sensitivity by several orders of magnitudes and have detected substantial numbers of faint sources; however, they have covered only a few percent of the sky.

Voges et al. (1999) identified X-ray sources of RASS-BSC with the Tycho Stars Catalog (Tycho-1; Hog et al. 1998) and then obtained the positional errors ($\sigma_{pos}$) from $6\arcsec$ to $75\arcsec$ with an average of $12\arcsec.5$ based on the spatial offsets between RASS-BSC sources and the Tycho catalog. A large search radius ($>30\arcsec$) is employed to match with the optical counterparts of Tycho-1 to the greatest extent possible. The matching radius is so large that contamination of the samples is inevitably introduced. However, because optically dim X-ray sources are invisible in optical catalogs, they could be assigned false counterparts. These imperfect identifications may lead to missing potential exciting candidates, such as isolated neutron stars (INSs).

%% INSs
X-ray sources with low optical luminosities, such as INSs (Neuhauser \& Trumper 1999; Treves et al. 2000; Rutledge et al. 2003), should be studied carefully. INSs are extremely optically dim with a relatively higher X-ray to optical flux ratio ($\lg(\ff) \sim 5.5$) (Treves et al. 2000). Considering the flux range of RASS, $B = 26 \sim 31$  mag for INSs is expected, which is below the detection limits for available large-scale optical sky surveys such as USNO (Monet et al. 2003) and the Sloan Digital Sky Survey (SDSS; York et al. 2000). Therefore, INS candidates hiding in RASS-BSC cannot be identified with optical counterparts in these large surveys. However, the $\sigma_{pos}$ of RASS-BSC require such a large search radius that it almost certainly covers optical counterparts brighter than $B=26$ mag. It is very difficult to identify INS candidates due to inadequate optical identifications. For example, the $\sigma_{pos}$ of INS candidate RX J0420.0-5022 given in BSC is $12\arcsec$, but the spatial offset between the BSC position and its infrared counterpart observed by \emph{Herschel} (Posselt et al. 2014) is larger than $16\arcsec.5$. Only seven INSs have been discovered to date (Treves et al. 2001; van Kerkwijk \& Kaplan 2007; Turolla 2009). The number of INSs in RASS-BSC that have been missed in previous studies due to poor positional accuracy and incorrect identifications should be investigated and evaluated.

%% Chandra:
The launch of the \emph{Chandra} Observatory in 1999 began a new era of X-ray astronomical research. This mission provides high-resolution X-ray spectra of various sources, X-ray stars and binaries, supernovae and their remnants, and interesting new populations (Paerels \& Kahn 2003). Identifications in multi-wavelengths of X-ray sources have been a basis for extensive astrophysical investigations (Ag\"{u}eros et al. 2009; Greiss et al. 2014; Parejko et al. 2008). \emph{Chandra} Advanced CCD Imaging Spectrometer (ACIS, Nousek et al. 1987) archival observations provide an X-ray source catalog with sharp positional accuracy. After more than 10 years of operation, \emph{Chandra} has observed a large number of RASS-BSC sources with somehow repeating frequencies. Utilizing the superb spatial resolution and sub-arcsec positional accuracy of \emph{Chandra} observations, the $\sigma_{pos}$ of BSC can be assessed and improved.

As the first of a series of articles, we cross-match X-ray sources between BSC and the ACIS catalog to obtain new spatial offsets for RASS-BSC. The INS candidates should be reconsidered with the better $\sigma_{pos}$ or search radius. In this article, we describe the poor spatial resolution and positional accuracy of RASS-BSC that can be assessed by matching with \emph{Chandra} ACIS sources. This paper emphasizes the direct assessment of the positional accuracy of BSC by comparison with their \emph{Chandra} positions.

The remainder of this article is organized as follows. The data and method that we adopt and the results are described in Section \ref{sec:data-method}. The discussion is presented in Section \ref{sec:dis}. A brief summary is presented in Section \ref{sec:summary}.

\section{Data and analysis}
\label{sec:data-method}

\subsection{\emph{ROSAT}}
Led by the German Aerospace Center and NASA, R\"{o}ntgensatellit (\emph{ROSAT}) observed all-sky soft X-ray sources during its mission from 1991 to 1999. With its PSPC, RASS is sensitive in the energy range of 0.1 to 2.4 keV (Tr\"{u}mper 1983). Voges et al. (1999) released RASS-BSC from the survey observations during the first half year (1990/1991) of the \emph{ROSAT} mission.

\begin{figure}
\centering
  \includegraphics[width=\textwidth, angle=0]{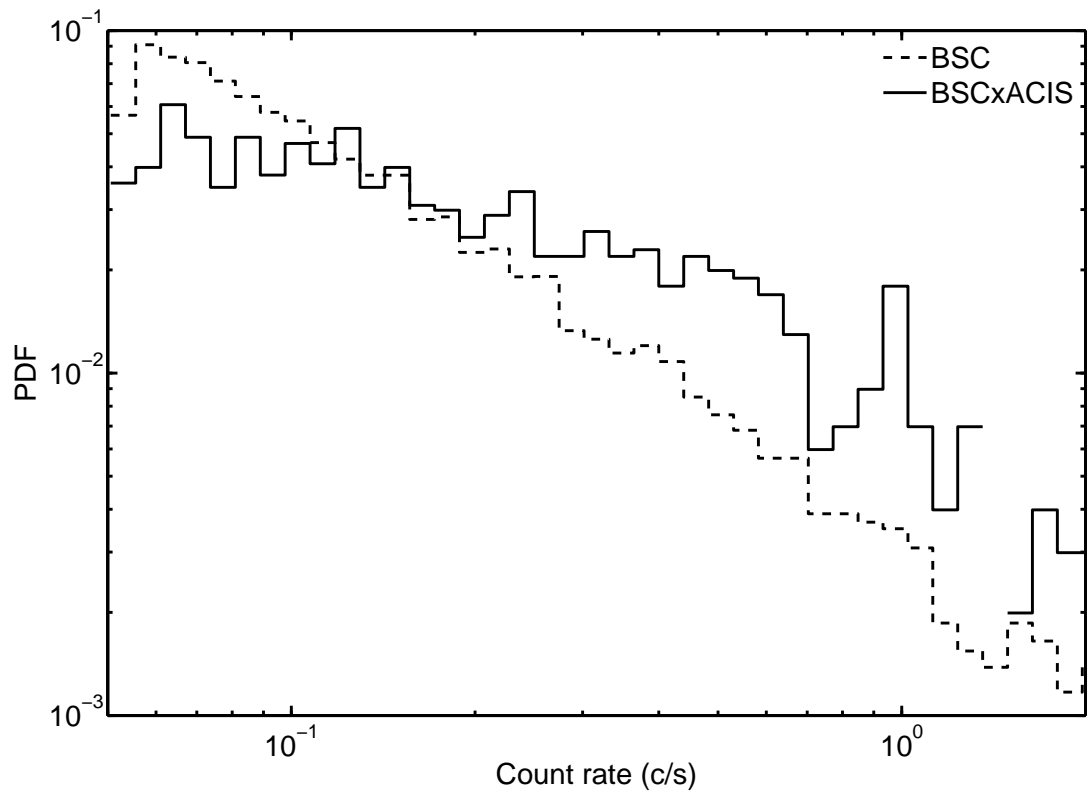}
  \caption{The distributions of count rates of 18,806 BSC and 1,004 BSCxACIS sources, which are both from BSC database. The logarithmic x- and y-axes are the count rate (in units of $c/s$) and normalized probability distribution function (PDF), respectively. The dashed and solid histograms represent entire BSC and BSCxACIS samples.}
    \label{fig:count}
\end{figure}

\begin{figure}
\centering
  \includegraphics[width=\textwidth, angle=0]{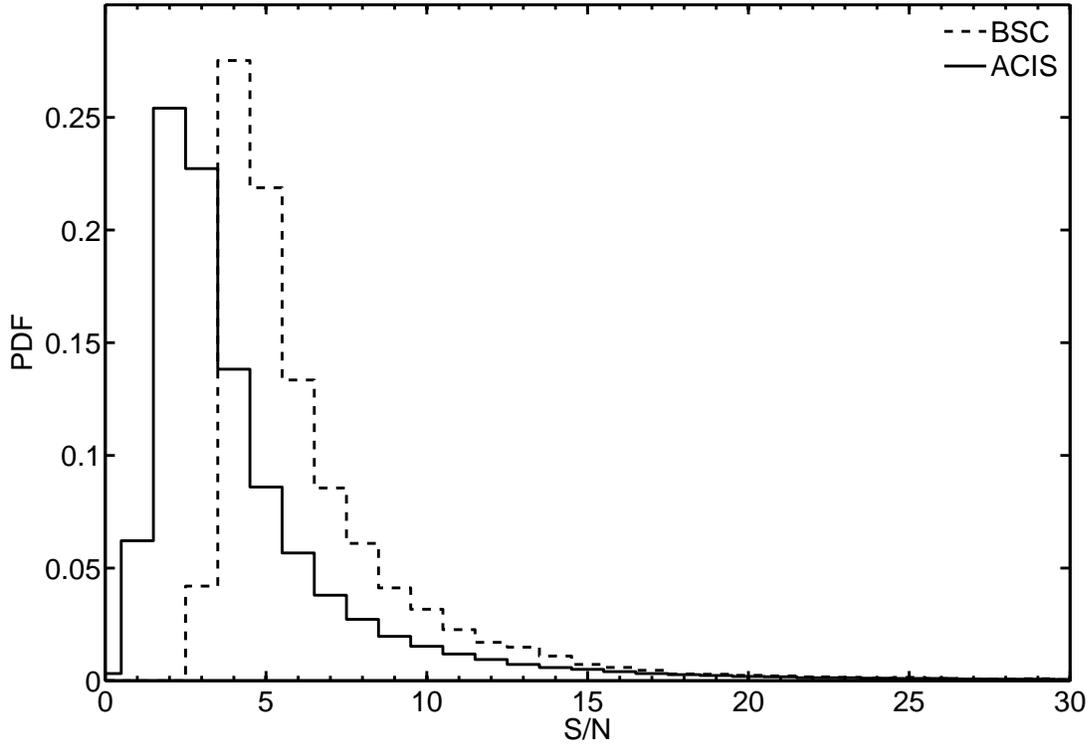}
  \caption{The distributions of observed signal-to-noise ratio (S/N) in BSC and ACIS. The S/N is limited within the range from 0 to 30 for convenient display. A few sources in BSC and ACIS with higher S/N are not covered. The y-axis is the normalized count binned by an interval of 1. The dashed and solid histogram represent BSC and ACIS, respectively.}
  \label{fig:snr}
\end{figure}

The BSC contains the spatial positions, count rates, exposure times, hardness ratios (HRs) and extent likelihood ($L_\mathrm{extent}$) of 18,811 sources. The distribution of BSC count rates is shown as a dashed histogram in Figure \ref{fig:count}. The signal-to-noise ratio (S/N) is derived as the ratio of the count rate to its uncertainty. The distribution of the S/N values of the BSC source is shown as a dashed histogram in Figure \ref{fig:snr}. The catalog provides two fluxes, which are determined by energy spectral distribution of AGNs and stars for each source. Utilizing the scanning mode of primary focus, RASS-BSC contains 18,811 X-ray sources with count rates that are larger than $0.05 s^{-1}$ in the 0.1---2.4 keV energy band, equivalent to a flux of $7\times 10^{-13} erg/s/cm^{-2}$ for AGNs or $3.75\times 10^{-13} erg/s/cm^{-2}$ for stars because different spectral distribution assumptions suggest inconsistent absolute luminosities. The typical exposure time of each scanning observation is from 300 to 600 seconds. The filed of view of ROSAT with the PSPC in the focal plane is about 2 degree diameter.

The $\sigma_{pos}$ of BSC sources are derived from the match between BSC and the other surveys, i.e. Tycho, Infrared Astronomical Satellite (IRAS), NASA Extragalactic Database (NED) etc. The minimum $\sigma_{pos}$ is $6\arcsec$ due to the systematic error. The maximum and median values of $\sigma_{pos}$ are $75\arcsec$ and $11\arcsec$, respectively.

BSC only contains bright sources, which distribute very uniformly in almost the entire sky.

\subsection{\emph{Chandra}}

The \emph{Chandra} ACIS is one of two focal plane instruments. This instrument is particularly useful because it can generate X-ray images while simultaneously measuring the energy of each incoming X-ray photon. The sizes of a CCD and a pixel of CCDs correspond to $8\arcmin.4$ and $0\arcsec.492$ on the sky, respectively (Garmire et al. 2003).

\begin{figure}
\centering
  \includegraphics[width=\textwidth, angle=0]{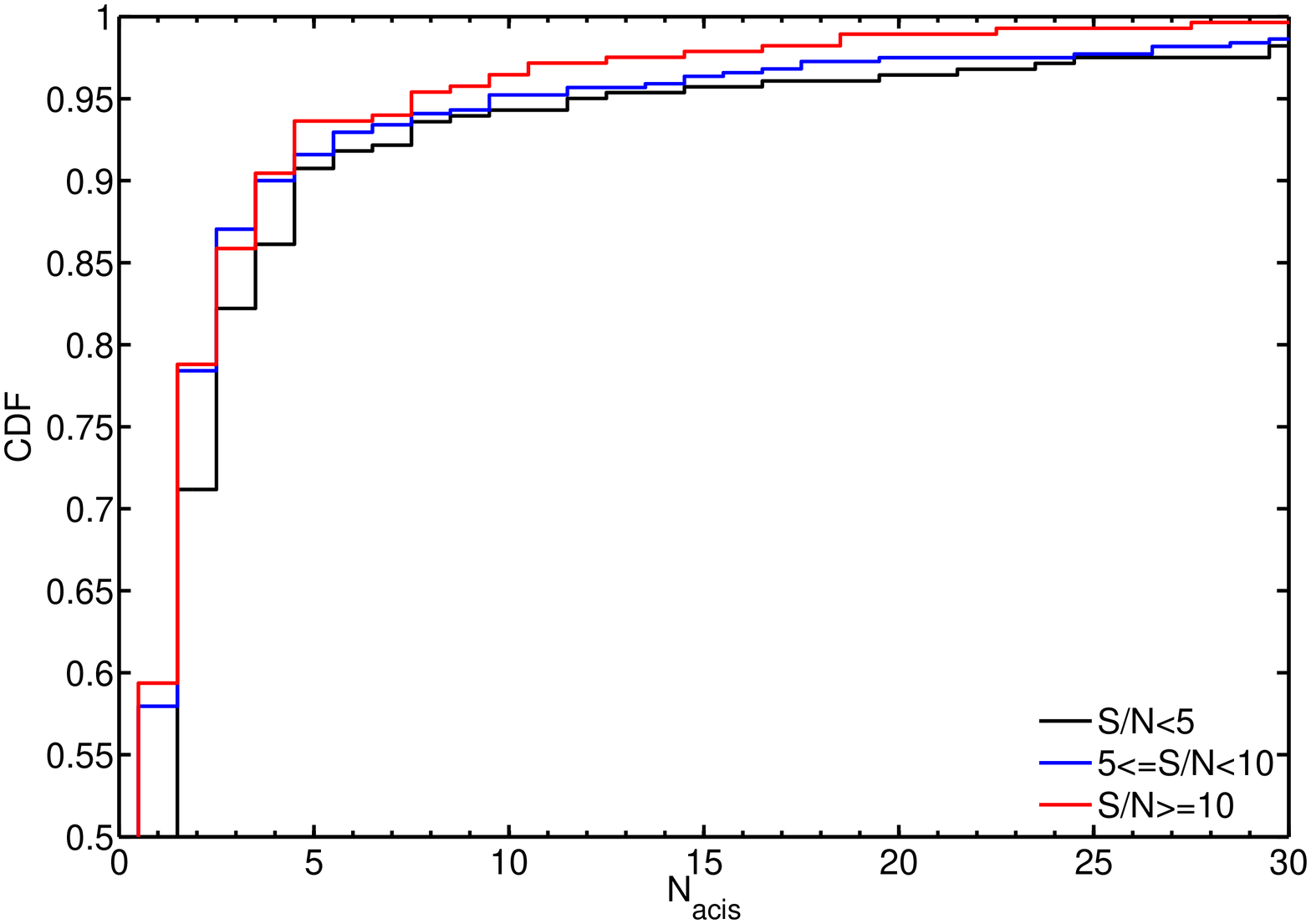}
\caption{The cumulative distribution function (CDF) of the counts of ACIS counterparts of BSCxACIS from 0 to 30. A longer tail is not shown. The three sub-samples divided by S/N intervals are plotted in different colors: black for $\mathrm{S/N}<5$, blue for $5\leq\mathrm{S/N}<10$ and red for $\mathrm{S/N}>10$.}
\label{fig:Nacis}
\end{figure}

For each ACIS observation, we apply the same procedures to detect and visually check point sources. The ACIS observations were downloaded from the \emph{Chandra} Data Archive on December 4, 2014. This leads to 10,029 ACIS observations, including 5,883 ACIS-I observations and 4,146 ACIS-S observations. For each observation, we use the on-axis chips, which include the S2 and S3 chips when the aimpoint is on S3 and all four I chips when the aimpoint is on an I chip. The ACIS-I and ACIS-S adopt four and two 1k x 1k pixel CCDs what has a pixel size of $0.492\arcsec$, corresponding to a field of view of about 280 and 140 arcmin$^2$, respectively. The exposure times cover a range from 50 s to 190 ks, a typical exposure value is about ten thousand seconds.

A wavelet detection algorithm, called {\tt wavdetect}, was used for point source detection; this algorithm is available in the \emph{Chandra} Interactive Analysis of Observations (CIAO, version 4.6) software package and is largely used for \emph{Chandra} observations (Freeman et al. 2012). We ran {\tt wavdetect} on each on-axis chip with scales of 1, 2, 4, and 8 pixels in the 0.3-8 keV band. The significance threshold was set to $10^{-6}$, equivalent to one potentially spurious pixel in one CCD. For the remaining parameters, we used the default values given in CIAO. The data analysis procedures led to the detection of 363,530 point sources.

Simulations by Kim et al. (2007) showed the positional error is usually less than $1^{\prime\prime}$ for a bright source, regardless of its off-axis angle (OAA); while for a weak source, it can increase to $4^{\prime\prime}$ at a large OAA (${\rm OAA} > 8^{\prime}$).

Each source record is then constructed by combining source detections in multiple observations (Wang et al. 2016), which leads to 217,828 distinct sources. For each source, when the individual detections are determined, the final position is computed by averaging the positions of individual detections with the detection significance as weights. The minimum positional error from individual detections is taken as the positional error.

\subsection{Method}

We employ the excellent positional accuracy of \emph{Chandra} ACIS to correct the $\sigma_{pos}$ of BSC sources. The nominal $\sigma_{pos}$ of BSC is only used to set a large search radius for each X-ray source. Our method is based on the following assumptions:

\begin{description}
  \item[-] Most \emph{Chandra} counterparts of BSC sources can be identified within the search radius of three times the nominal error ($3\sigma_{pos}$) of BSC sources if the footprint of BSC sources had been covered by \emph{Chandra}.
  \item[-] Clustered multiple sources observed by the highly resolved \emph{Chandra} view are perhaps regarded as single sources with extent in BSC (see extent likelihood of BSC for details).
  \item[-] During the time gap between \emph{ROSAT} and \emph{Chandra}     observations, dramatic decreases in X-ray luminosities are rare. No source of BSC disappears in the same ACIS fields of view.
  \item[-] The astrometry is reasonably good for both \emph{ROSAT} and \emph{Chandra} observations so that there is no astrometric offset between observations.
\end{description}

To locate the BSC sources on \emph{Chandra} ACIS images, for each BSC source (with observed position), we rank the ACIS sources within a $3\sigma_{pos}$ circle centered on a BSC position by X-ray fluxes. We consider the average position of ACIS sources weighted by their fluxes to be an equivalent BSC position. The offset between the claimed and derived equivalent position is the credible positional error $\sigma'_{pos}$. If the ranked ACIS sources are dominated by the strongest one, then the derived equivalent position is close to that.

The derived average position is located inside the clustering area ($3\sigma_{pos}$ circle). We do not replace the nominal BSC position with this average position because the average center doesn't represent real ACIS source. ROSAT with lower spatial resolution mistakes multiple sources for single sources. Each ranked ACIS source should be utilized to investigate the optical counterparts of nominal BSC sources. Each position of ACIS source is matched with optical catalogs.

\begin{figure}
\centering
 \includegraphics[width=0.4\textwidth, angle=0]{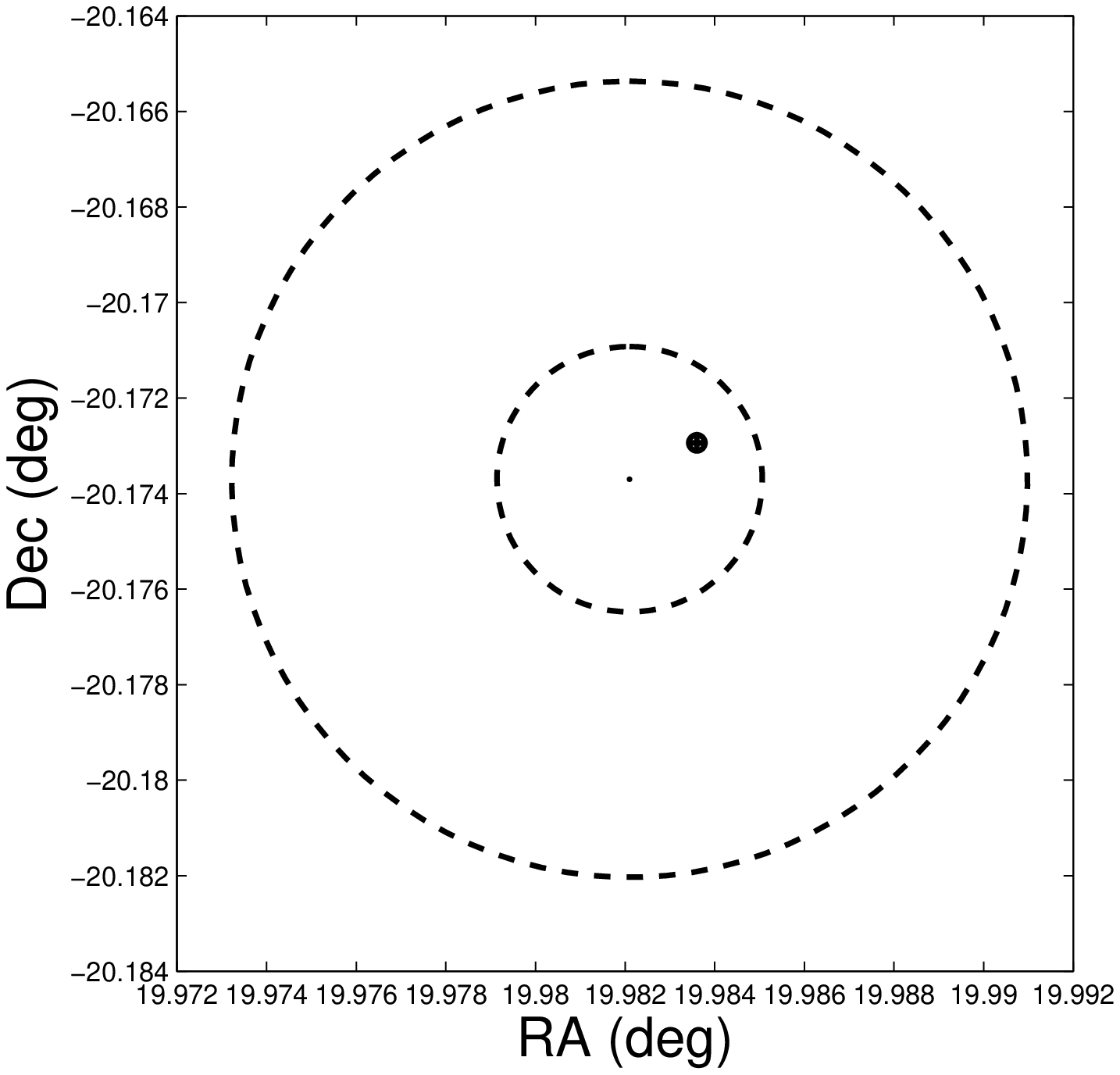}
  \includegraphics[width=0.42\textwidth, angle=0]{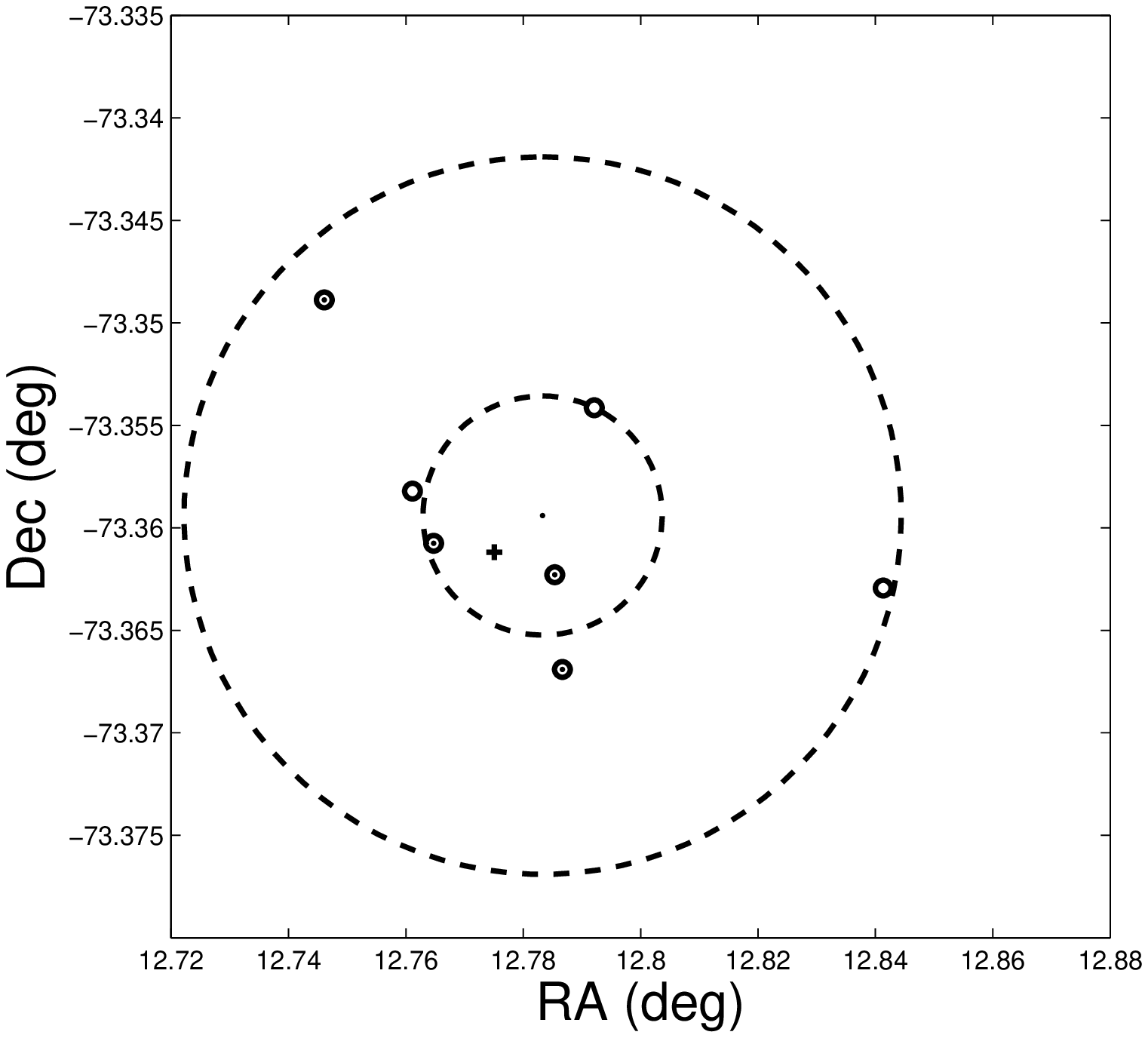}
\caption{Two cases of our matching method. The large dashed curves are   $\sigma_{pos}$ and $3\sigma_{pos}$ circles. Left panel: The figure   shows the BSCxACIS source with only one counterpart ACIS source   within the $3\sigma_{pos}$ circle. Right panel: There are 7 ACIS   sources within the $3\sigma_{pos}$ circle. Among them, dominated   sources (flux weights $> 10\%$) are indicated as open circles with   centered dots. The weighted average position is labeled with a plus   symbol "+''. The distance between the plus symbol and the center of   this plot is the derived positional error. Sources out of the $3\sigma_{pos}$ circles are not shown. }
  \label{fig:2cases}
\end{figure}

We consider the corresponding ranked ACIS sources as our cross-matching sample, which is named as BSCxACIS. This sample contains 1,004 BSC sources and the corresponding 3,487 \emph{Chandra} sources, which is approximately $5\%$ of the entire BSC records and does not show any preference in terms of the direction in the sky. The cumulative distribution function (CDF) of the number of corresponding ACIS sources of each BSCxACIS source is shown in Figure \ref{fig:Nacis}. For instance, two cases of 1,004 results are shown in Figure \ref{fig:2cases}. Further discussions and results are presented in Section \ref{sec:pd}.

\subsection{Positional error}
\label{sec:pd}

Within the search radius of three times the nominal positional errors $3\sigma_{pos}$, ACIS sources are selected and are considered to be potential counterparts of BSC. Compared with the BSC observed position, the average position of these counterparts with weights of X-ray fluxes helps provide the improvement of the positional accuracy of each BSC source in the BSCxACIS sample.

\begin{figure}
\centering
 \includegraphics[width=\textwidth, angle=0]{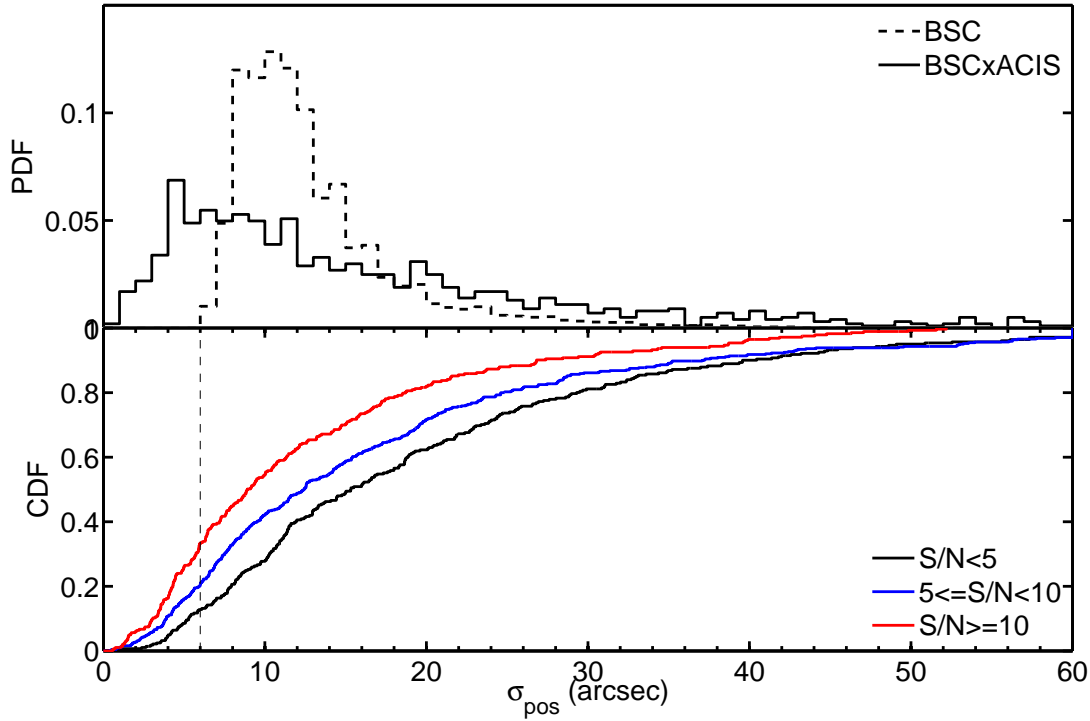}
\caption{The distributions of spatial positional errors of BSC (dashed   histogram) and BSCxACIS (solid histogram). The x-axis is the positional errors $\sigma_{pos}$ in units of arcsec. Very few sources are located out of the range shown in the figure ($>60\arcsec$). The y-axis is the normalized PDF (upper panel) and CDF (lower panel). The peaks of distributions are approximately $11\arcsec$ and $4\arcsec$ for BSC and BSCxACIS, respectively. The BSCxACIS sample is divided into three S/N catalogs that are represented by different colors (color code is similar to that in Figure \ref{fig:Nacis}). The vertical dashed line shows the position of $6\arcsec$ that is the lower limit of original claimed positional error. }\label{fig:poserr}
\end{figure}

As shown in Figure \ref{fig:count}, the count rate distribution of BSCxACIS is slightly flatter than that of the original entire BSC sample. BSCxACIS includes less faint sources and a higher fraction of strong sources. The possible reasons are detected limit of Chandra is brighter than BSC and a fraction of transit sources in BSC disappeared in ACIS.

\begin{figure}
\centering
 \includegraphics[width=\textwidth, angle=0]{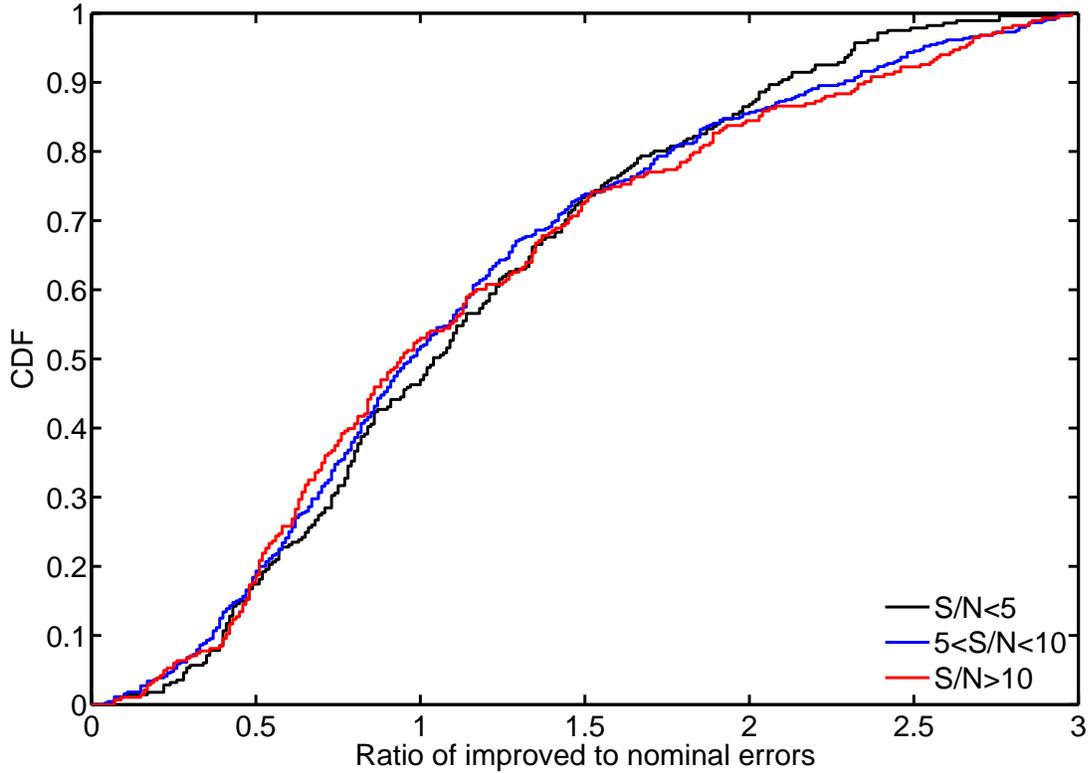}
\caption{The CDF distribution of spatial positional errors $\sigma'_{pos}$ of BSCxACIS in units of nominal errors $\sigma_{pos}$. The BSCxACIS sample is divided into three S/N intervals that are represented by different colors (color code is the same as that in Figure \ref{fig:Nacis}).}\label{fig:sig-sig}
\end{figure}

Half (561 sources) of BSCxACIS refers to only one ACIS source, whereas the other 443 BSC sources are considered to be multiple targets of \emph{Chandra} ACIS. The number distribution (CDF) is shown in Figure \ref{fig:Nacis}. The BSCxACIS sample is divided into three catalogs with different S/N intervals ($\mathrm{S/N}<5$, $5\leq\mathrm{S/N}<10$, and $\mathrm{S/N}\geq 10$). The number of \emph{Chandra} counterparts do not show any relationship with the S/N. For the source with a single ACIS counterpart, the spatial offset between it and the BSC position is the new and true positional error $\sigma'_{pos}$. For multiple \emph{Chandra} ACIS ranked targets, the sum of the weights of ranked fluxes is normalized to 1. We define dominated sources as the sources with weights larger than $10\%$. In Figure \ref{fig:2cases}, the left panel shows a case of a single target, and the right panel shows a case of multiple counterparts. The distributions of all $\sigma_{pos}$ and $\sigma'_{pos}$ are compared in Figure \ref{fig:poserr} and Figure \ref{fig:sig-sig}. A considerably higher fraction  of small $\sigma'_{pos}$ is collected. Half (508 sources) of BSCxACIS obtains smaller positional errors. However, our method generates a number samples with relatively larger $\sigma'_{pos}$. The positional errors of 21 sources exceed $60''$, while only 2 sources have nominal $\sigma_{pos}>60''$ that are not shown in the range of Figure \ref{fig:poserr}. No sources are with $\sigma_{pos}<6''$ due to minimum systematic errors, which contrasts sharply with a very small peak of the $\sigma'_{pos}$ profile, i.e., $\sim 4''$. To provide a clearer comparison between the nominal $\sigma_{pos}$ and new corrected $\sigma'_{pos}$, we define a ratio of $\sigma'_{pos}$ to $\sigma_{pos}$ and plot its distribution in Figure \ref{fig:sig-sig}.

We observe that a fraction ($49\%$) of the BSCxACIS sample has larger positional errors derived by the \emph{Chandra} ACIS counterparts. We consider the relationship between the number of ACIS counterparts $N_\mathrm{ACIS}$ and the ratio $\sigma'/\sigma$ of these sub-samples. However, no clear trend is found.

Because nominal positional errors cannot be smaller than $6''$, we count the fraction of sources with $\sigma'_{pos}<6''$ and consider the relationship between the fraction and signal-to-noise ratio $\mathrm{S/N}$ of the BSC observation. In the lower panel of Figure \ref{fig:poserr}, the left edge of the dashed histogram is the limit of $6''$, which does not bound the $\sigma'_{pos}$ of BSCxACIS (solid histogram). From the results shown in Figure \ref{fig:poserr}, we learn that a higher $\mathrm{S/N}$ generates more accurate position measurements for true results.

We can analyze the lumped spatial distribution of BSCxACIS X-ray sources with the help of the ranked ACIS fluxes. The flux weights $w^{(i)} \,(i=1, 2, \ldots, N_\mathrm{ACIS}, \sum_1^{N_\mathrm{ACIS}} w^{(i)}=1)$ of the counterparts of each BSCxACIS source are an important indicator for describing the domination of clustering X-ray sources. The count of $w^{(i)}>\%$ of each BSCxACIS source is listed in the second row of Table \ref{tab:cluster}. The third row is the count of $w^{(i)}>20\%$ and so on. The sum of each row of Table \ref{tab:cluster} is 1,004. The empty elements correspond to zero.

\begin{table}[htb]
\caption{Number of different flux weights.
For one BSCxACIS source, each ACIS counterpart has a different flux weight $w^{(i)}$ and $\sum_1^{N_\mathrm{ACIS}} w^{(i)}=100\%$. The second row presents the distribution of $w^{(i)}>10\%$. Among the 1,004 BSCxACIS sources, only one source has all counterparts with $w^{(i)}$ less than $10\%$, while 739 BSCxACIS sources have 1 counterpart with $w^{(i)}$ larger than $10\%$, 172 sources have 2 counterparts with weights larger than $10\%$, and so on. The remaining rows provide higher weight constraints. All empty elements are zero. The sum of any row is 1,004.
}\label{tab:cluster}
  \begin{tabular}{c|rrrrrr}
    \hline
    \hline
                           & 0  &  1  & 2   & 3  & 4  & 5  \\
    \hline
    $w^{(i)}>10\%$ & 1  & 739 & 172 & 69 & 18 & 5  \\
    $w^{(i)}>20\%$ & 11 & 818 & 155 & 20 &    &    \\
    $w^{(i)}>30\%$ & 22 & 877 & 102 & 3  &    &    \\
    $w^{(i)}>40\%$ & 53 & 912 & 39  &    &    &    \\
    $w^{(i)}>50\%$ & 85 & 919 &     &    &    &    \\
    $w^{(i)}>60\%$ & 154& 850 &     &    &    &    \\
    $w^{(i)}>70\%$ & 203& 801 &     &    &    &    \\
    $w^{(i)}>80\%$ & 239& 765 &     &    &    &    \\
    $w^{(i)}>90\%$ & 280& 724 &     &    &    &    \\
    \hline
  \end{tabular}

\end{table}

Because the spatial resolution of \emph{ROSAT} is lower than that of \emph{Chandra} or most optical surveys, BSC perhaps treats clustered X-ray sources as single targets (point or extended). A fraction of the sources belong to clusters, but they are identified as single sources by \emph{ROSAT}. We calculate the accuracy rate of point sources within different $S/N$ bins. As listed in Table \ref{tab:cluster2}, the accuracy rate is from approximately $48\%$ to $54\%$ when $S/N$ varies. Along with increasing $S/N$, the false rate stabilizes at approximately $50\%$. This result implies that even if we exclude effects due to observational error, the counts of point sources are still overestimated by $50\%$. These false ``point sources'' can be resolved by \emph{Chandra} observations.

\begin{table}[htb]
\caption{Number and accuracy rate of point sources. The right three columns list the results of different S/N intervals. The second row is the total number of BSCxACIS samples. The third row presents the number of nominal ``point sources'' defined by $Lextent=0$  in BSC. The fourth row presents the true number of point counterparts among the nominal point sources. Finally, the last row lists the accuracy rate, i.e., $N_\mathrm{point}/N_\mathrm{Lextent=0}$.}\label{tab:cluster2}
  \begin{tabular}{cccc}
    \hline
    \hline
                   & $S/N<5$ & $5\leq S/N<10$    & $S/N\geq10$ \\
    \hline
    $N_\mathrm{BSCxACIS}$ &   281   &  440          &    283      \\
$N_\mathrm{Lextent=0}$  &   149   &  132         &      23     \\
true $N_\mathrm{point}$   &    81   &   71         &      11     \\
    accuracy rate  & $54.4\%$&  $53.8\%$    &  $47.8\%$  \\
    \hline
  \end{tabular}

\end{table}

\begin{figure}
\centering
 \includegraphics[width=\textwidth, angle=0]{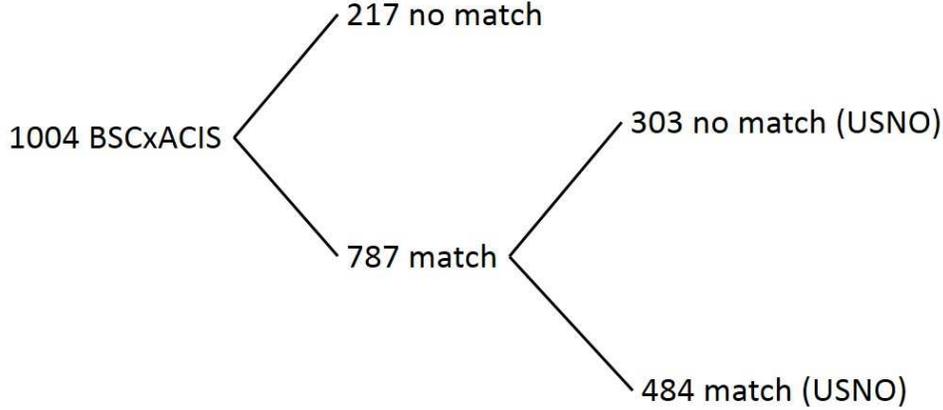}
\caption{The schema of more potential INS candidates using corrected   positional errors. Only 217 of 1,004 BSCxACIS sources cannot be matched by the USNO catalog with nominal errors. If the ``matched'' 787 sources are processed using the new derived positional errors, 303 more non-identifications are revealed. }\label{fig:ins}
\end{figure}

Based on the BSCxACIS sample, we obtain the following facts:

\begin{description}
  \item[-] In the coverage area of ACIS images, \emph{Chandra} counterparts related with 1,004 BSC sources can be found within the search radius of $3\sigma_{pos}$. For each BSCxACIS source, the number of \emph{Chandra} counterparts is between 1 to more than 30.
  \item[-] For 1,004 BSCxACIS sources, each one corresponds to an ACIS X-ray source or a set of ACIS X-ray sources with different X-ray fluxes. With weights calculated from fluxes, an average position is determined for each BSCxACIS source. We consider the offset between the claimed and derived positions as the corrected positional error $\sigma'_{pos}$.
  \item[-] The peak of the $\sigma'_{pos}$ distribution is     approximately $4''$. The $\sigma'_{pos}$ of 508 BSCxACIS sources     are smaller than their original nominal values $\sigma_{pos}$. The     fraction of sources with $\sigma'_{pos}<6''$ to all BSCxACIS     samples increases along with the $\mathrm{S/N}$ of BSC observations (see the lower panel of Figure \ref{fig:poserr}), while all nominal $\sigma_{pos}$ are larger than $6''$.
  \item[-] A fraction of the BSC sources are considered to be X-ray     clusters. A total of 264 BSCxACIS sources have more than one counterpart with flux weights larger than $10\%$, and 175 sources have more than one counterpart with flux weights larger than $20\%$.
\end{description}

\section{Discussion}
% INS
\label{sec:dis}

Due to the combined effects of positioning accuracy and spatial resolution, mis-identifications of dim optical counterparts lose several interesting objects in the Milky Way, such as INSs.

ACIS counterparts of five INSs are found within $3\sigma_{pos}$
circles (the first five rows of Table \ref{tab:ins}), and only one
ACIS source exists for each INS in BSC. Their positional errors (the
third and fourth columns of Table \ref{tab:ins}) are updated by ACIS
sources. The other two INSs have not been covered by the footprint of \emph{Chandra} ACIS.

\begin{table*}[htb]
\caption{BSC and BSCxACIS properties of seven known INSs that were found in BSC previously. The first column lists the name of the \emph{ROSAT} X-ray source. The second column presents the number of ACIS counterparts of each BSC source. The third and fourth columns list the nominal and newly derived positional errors in units of arc seconds. The fifth column presents the count rate (in units of $s^{-1}$) of X-rays observed by \emph{ROSAT} and listed in BSC. The sixth column presents the optical photometric constraints of each INS. Finally, the last column lists references.}\label{tab:ins}
  \begin{tabular}{lrrrrrcc}
  \hline
  \hline
Object & $N_{ACIS}$ & $\sigma_{pos}$ & $\sigma'_{pos}$ & Count Rate & Optical & Ref. \\
       &            & $''$           & $''$            & $s^{-1}$ &  mag & \\
  \hline
  RX J0420.0-5022   & 1 & 12 & 16.13  & 0.12  & $B=26.6$ & 1 \\
  RX J0720.4-3125   & 1 & 7  & 4.05   & 1.70  & $B=26.6$ & 2 \\
  RX J0806.4-4123   & 1 & 8  & 4.98   & 0.33  & $B>24$   & 1 \\
  RBS 1223          & 1 & 7  & 5.94   & 0.29  & $m_{50ccd}=28.6$ & 3 \\
  RX J1605.3+3249   & 1 & 7  & 10.43  & 0.08  & $B=27.2$ & 4  \\
  RX J1856.5-3754   & 0 & 7  & -      &  3.60 & $V=25.7$ & 5 \\
  RBS 1774          & 0 & 9  & -      &  0.23 & $R>23$   & 6 \\
  \hline
\end{tabular}

References:
(1) Haberl et al. (2004) (2) Haberl et al. (1997) (3) Haberl et al. (2003) (4) Motch et al. (2005) (5) Burwitz et al. (2003) (6) Zane et al. (2005)

\end{table*}

\begin{figure}
\centering
 \includegraphics[width=\textwidth, angle=0]{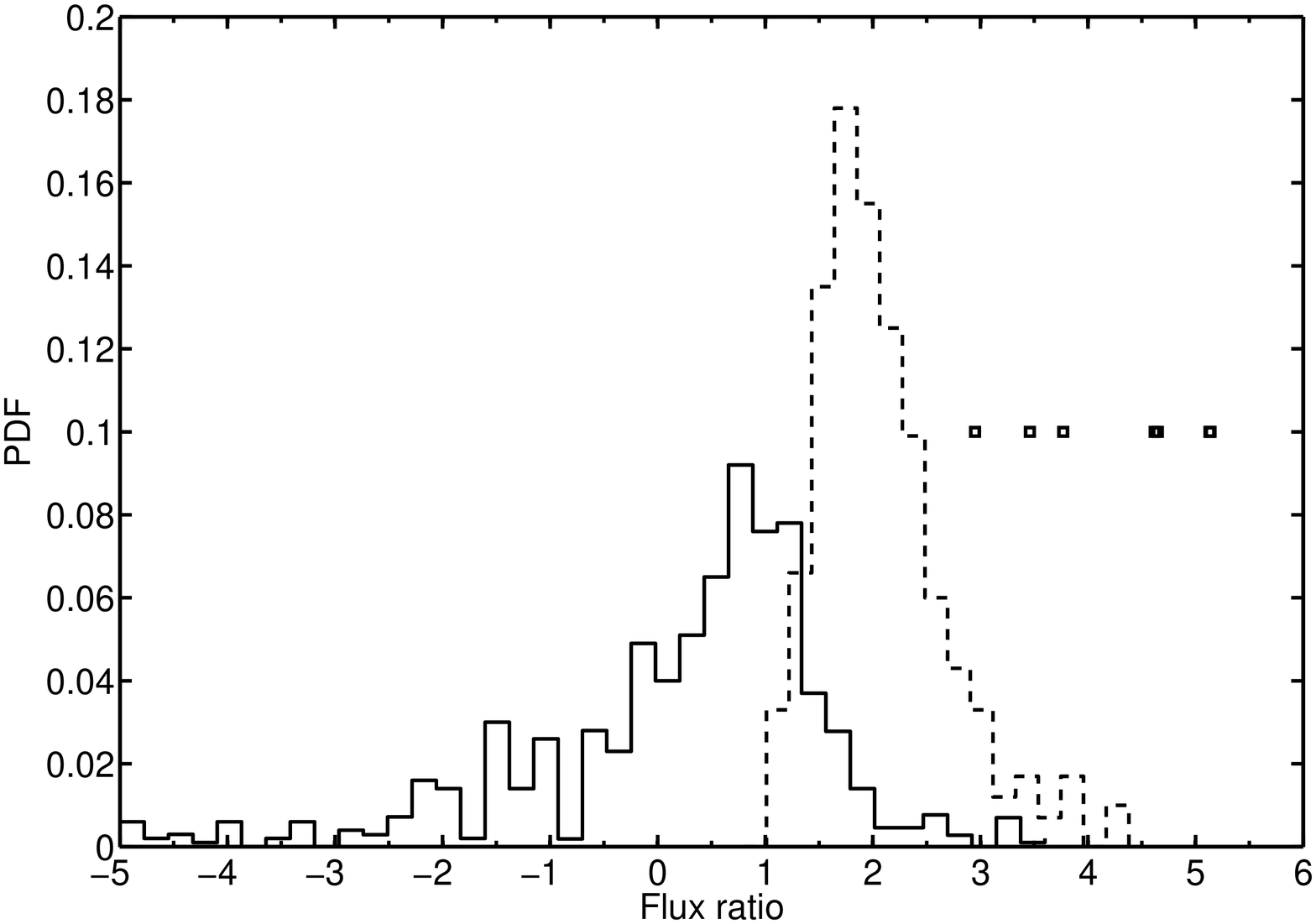}
\caption{The logarithmic flux ratio distribution of BSCxACIS sources. The solid line is the distribution of sources with USNO B1.0 optical counterparts. The dashed line is the distribution of the lower limits of sources that cannot be identified with the USNO B1.0 catalog. The open squares are the flux ratios of a few known INSs.}\label{fig:insf}
  \end{figure}

The determination of INS candidates requires a series of multi-band observations and related detailed analysis, which depends on longer period observations or sky surveys. However, the first criterion of INSs is optical invisibility of the X-ray source for almost all imaging surveys. The inappropriate positioning and positional errors (e.g., a too large search radius) of BSC sources perhaps leads to INS candidates being assigned to ``match'' incorrect optical counterparts. According to the number density estimation at a moderate location ($l=28^\circ$, $b=50^\circ$), within a search radius of $30''$, one can always find 6.7 stars at the limiting magnitude $g=25$ for SDSS and even 0.7 star within a search radius of $10''$. Considering the energy band of \emph{ROSAT}, Treves et al. (2000) estimated $\lg{(f_X/f_{V})}\sim 5.5$ for INSs, which suggests that the optical photometry of INSs should not be brighter than $26$ mag. The almost certain presence of contaminating objects brighter than $B=26$ mag is due to the too large search radius and the low spatial resolution of \emph{ROSAT}, which makes it extremely difficult to identify INS candidates.

The positions from BSC and ACIS are used to match with the USNO B2.0 survey data. Among the 1,004 BSCxACIS sources, even though we use the new positional errors, 217 sources cannot be matched with optical counterparts from the USNO B2.0 catalog by their BSC positions and search radius. The remaining 787 sources can match their optical counterparts of the USNO B2.0 catalog by their BSC positions within a $3\sigma$ error circle. We match these 787 sources with the USNO B2.0 catalog with considerably more accurate ACIS positions. Among them, the new matching process reports that the 303 sources cannot obtain optical matching by ACIS positions.

If we correct the positions of BSC with ACIS, we will obtain 303 more sources without optical counterparts (see the logic sequence in Figure \ref{fig:ins}). The 303 sources cannot be identified by apparent USNO optical counterparts with corrected positions and errors, but they have sufficiently strong X-ray emission, whereas only $217$ candidates are collected by using the original positions and errors. The improvement of positions increases almost \textbf{$1.5$ times} the non-identification sources. Of course, true INSs are only a subset of these non-identification sources, but increased samples help find back sources that are missed by incorrect positions. For the entire BSC sample, as we know, the original positions finally provide 7 INSs. Along this proportion, if the sample distribution does not depend on any a priori trend, the new positioning result would reveal \textbf{$\sim 10$ more} potential INS candidates, although these candidates still need to be verified by more detailed criteria and observations.

\section{Summary}
\label{sec:summary}

The RASS conducted during the first half year of the mission represents by far the most complete soft X-ray sky survey. \emph{Chandra} ACIS data provided supplementary superb position information and better spatial resolution, which can be used to review and correct the positions of bright sources observed by ROSAT. For this goal, we assessed the positional errors for BSC sources by direct comparison with their \emph{Chandra} positions. We evaluated the nominal positional errors quoted in BSC based on the correlation between Tycho stars and BSC sources. We corrected the positions and positional errors of a fraction of \emph{ROSAT} BSC X-ray sources by directly utilizing the excellent positioning of \emph{Chandra} ACIS observations. An ACIS catalog has been compiled from images of \emph{Chandra} ACIS, which contains multi-observations of the same sources.

We discussed the new positional errors of BSCxACIS samples with S/N intervals and number of ACIS counterparts. The accuracy rate of nominal ``point sources'' was estimated by comparison with weights of multi-counterparts within the search radius of BSC sources. The count rates, S/N, clustering and point-source likeness of BSC sources were analyzed using the new derived positional errors and S/N intervals. Many investigators have achieved excellent results with the \emph{ROSAT} mission and its BSC release, such as seven INSs. Based on this great and important survey of X-rays, a slight improvement may provide the opportunity to find more interesting results.

To evaluate the missing INS candidates, we matched BSCxACIS with the USNO B1.0 catalog by Chandra positions. Ten additional INS candidates might be revealed in the future due to improvement of positional error and follow-up observations of RASS X-ray sources.

In the near future, the faint source catalog (FSC) of \emph{ROSAT} will be used to compare with the other optical surveys and ACIS. New results can be expected. Additionally, we expect to investigate the source variability both in the flux and in the hardness ratio through our subsequent articles. The temporal baseline from less than 1 year to more than 10 years during the gap between \emph{ROSAT} and \emph{Chandra}, between BSC and ACIS observations, is sufficient for comparing variability behaviors for sources of different types. In the second article of this series, we will discuss the variability of BSCxACIS sources during \emph{ROSAT} to \emph{Chandra} observations. The combined time baseline of BSCxACIS observations allows us to probe a statistically significant number of X-ray sources.

\begin{acknowledgements}
This work is supported by the National Natural Science Foundation of China (NSFC) No. 11503002 and 11533002, and ``the Fundamental Research Funds for the Central Universities''. We have made use of the \emph{ROSAT} Data Archive of the Max-Planck-Institut f\"{u}r extraterrestrische Physik (MPE) at Garching, Germany. The scientific results reported in this article are based on data obtained from the \emph{Chandra} Data Archive.
\end{acknowledgements}

\end{document}